# Network connection strengths: Another power-law?

Chunguang Li [1], Guanrong Chen [2,*]

1. College of Electronic Engineering, University of Electronic Science and Technology of China, Chengdu, 610054, P. R. China

2. Department of Electronic Engineering, City University of Hong Kong, 83 Tat Chee Avenue, Kowloon, Hong Kong, P. R. China

*Corresponding author, Email: gchen@ee.cityu.edu.hk   Phone: (852) 2788 7922
Fax: (852) 2788 7791

**Abstract**: It has been discovered recently that many social, biological and ecological systems have the so-called small-world and scale-free features, which has provoked new research interest in the studies of various complex networks. Yet, most network models studied thus far are binary, with the linking strengths being either 0 or 1, while which are best described by weighted-linking networks, in which the vertices interact with each other with varying strengths. Here we found that the distribution of connection strengths of scientific collaboration networks decays also in a power-law form and we conjecture that all weighted-linking networks of this type follow the same distribution.

The finding of small-world[1] and scale-free[2] properties has provoked new research interest in the studies of complex networks[3-8]. Most network models studied thus far are binary, with the linking strengths being either 0 or 1. The main concern in studying these network models was the topological namely structural properties of the networks. However, many complex networks are best described by weighted-linking networks, in which the vertices interact with each other with varying strengths. There are also some studies of weighted networks[9-10], but the approaches were to subjectively "assign" different weights to the links connecting vertices..

Here, we study statistical properties of the connection strengths of "real-world" networks by analysing three examples of scientific collaboration networks, in which the vertices are authors and a connection strength is defined as the number of collaborations (joint publications) between a pair of authors. The topological structures of these networks are different: the first one is a star-shaped network, and the second and the third are small-world networks.

## 1. Connection strengths of scientific collaborate networks

We study the statistical properties of the connection strengths of three examples of scientific collaboration networks in this section.

The first example is the Erdös's collaboration network[11], which is a star-shaped network with Erdös at the centre. Paul Erdös, a great mathematician who dies on September 20, 1997, at the age of 83, published more than 1,600 mathematical research papers in his lifetime. He had 507 coauthors, in which there were 306 coauthors who published 1 paper with him, and in which the coauthor having the largest number of joint publications with him was Sarkozy, 62 papers. Figure 1 shows the distribution, $P(k)$, of the connection strengths $k$ (number of joint publications) between Erdös and his coauthors, in a double-logarithmic plot. The data plot follows a power-law distribution; that is, $P(k) = k^{-\alpha}$, with an exponent $\alpha = 2.0 \pm 0.2$. The straight lines in the double-logarithmic plots have slopes $\alpha = 2.0$, corresponding to the exponents in the power-law distributions.

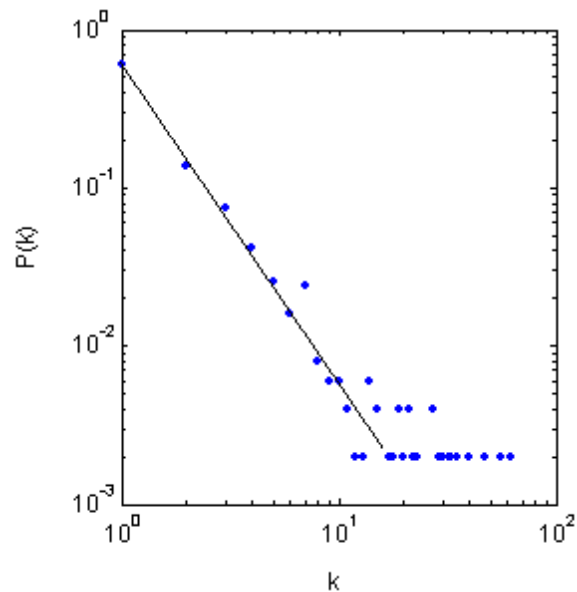

Fig.1 The connection strengths distribution of the Erdös collaboration network.

The second example is the coauthorship network in the field of chaos control and synchronization[12] during the time period of 1987-1997. After excluding papers of single authors (they do not contribute to the coauthorship network), we obtain a network of 740 vertices. The probability distribution, $P(k)$, of the connection strengths, defined as above, is shown in Figure 2. The data plot also closely follows a power-law in a double-logarithmic plot, with an exponent $\alpha = 2.5 \pm 0.1$. The straight lines in the double-logarithmic plots have slopes $\alpha = 2.5$, corresponding to the exponents in the power-law distributions.

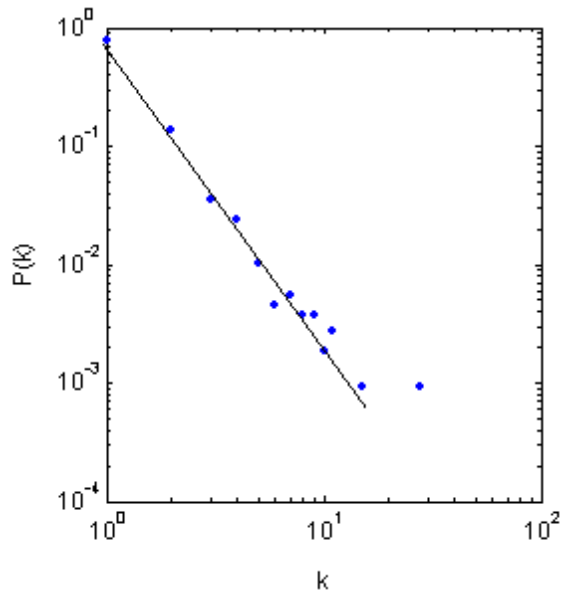

Fig.2 The connection strengths distribution of the coauthorship network in the field of chaos control and synchronization.

The third example is the mollusk research[13] coauthorship network. We use the data in the time period of 1990-2001. There are a total of 4196 authors, after excluding papers of single authors. In this case, the data plot obeys a rather perfect power-law distribution, with an exponent $\alpha = 3.5 \pm 0.1$, as shown by Figure 3 in double-logarithmic plot. The straight lines in the double- logarithmic plots have slopes $\alpha = 3.5$, corresponding to the exponents in the power-law distributions.

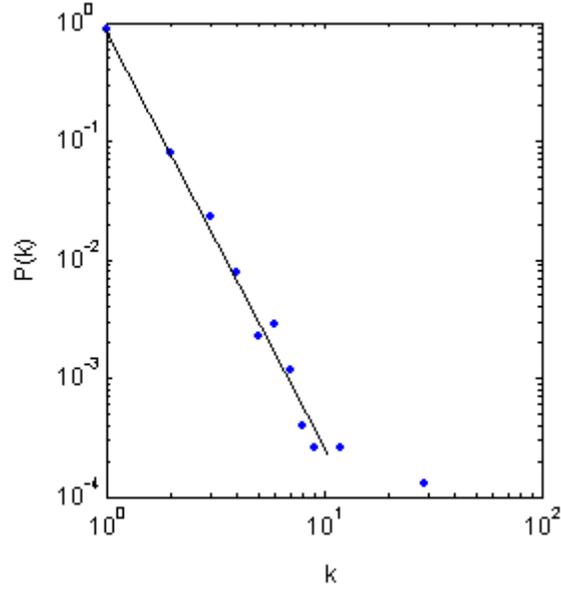

Fig. 3 The connection strengths distribution of the mollusk research coauthorship network.

## 2. Discussion and conjecture

The most interesting and important finding is the power-law nature of the *connection strengths* distribution of scientific collaboration networks. It is another power-law in complex networks, besides the power-law *vertex degree* distributions of scale-free networks studied in the current literature[2]. Some other social networks, such as the film-actors collaboration network and the company-directors relation network, are similar to the above-discussed scientific collaboration networks. The connection strengths in these networks should also follow the same power-law distribution. We conjecture that the connection strengths of all weighted-linking networks of this type, such as biological neural networks, electronic circuits, language networks, and various social networks, also possess this power-law distribution.

The discovery of the power-law degree distributions (as well as the preferential attachment) of scale-free networks has important implications for a host of applications, from drug development to Internet security [14], so does our finding of the power-law distribution of the connection strengths of complex networks.